\documentclass[prd,amsmath,nofootinbib,amssymb,superscriptaddress,preprintnumbers,twocolumn,10pt]{revtex4}

\pdfoutput=1

\usepackage{graphicx}
\usepackage{dcolumn}
\usepackage{bm}
\usepackage{amssymb}
\usepackage{latexsym}
\usepackage{booktabs}
\usepackage{amsmath}
\usepackage{multirow}
\usepackage{url}

\usepackage{float}
\usepackage[colorlinks=true, linkcolor=red, citecolor=blue]{hyperref}
\usepackage{hyperref}

\newcommand{\ve}{\varepsilon}

\usepackage[normalem]{ulem}
\usepackage{color}
\usepackage{array}
\usepackage{enumerate}

\begin{document}

\title{Impacts of gravitational-wave standard siren observations from Einstein Telescope and Cosmic Explorer on weighing neutrinos in interacting dark energy models}

\author{Shang-Jie Jin}
\affiliation{Department of Physics, College of Sciences, Northeastern University, Shenyang 110819, China}
\author{Rui-Qi Zhu}
\affiliation{Department of Physics, College of Sciences, Northeastern University, Shenyang 110819, China}
\author{Ling-Feng Wang}
\affiliation{Department of Physics, College of Sciences, Northeastern University, Shenyang 110819, China}
\author{Hai-Li Li}
\affiliation{Department of Physics, College of Sciences, Northeastern University, Shenyang 110819, China}
\author{Jing-Fei Zhang}
\affiliation{Department of Physics, College of Sciences, Northeastern University, Shenyang 110819, China}
\author{Xin Zhang\footnote{Corresponding author}}
\email{zhangxin@mail.neu.edu.cn}
\affiliation{Department of Physics, College of Sciences, Northeastern University, Shenyang 110819, China}
\affiliation{Frontiers Science Center for Industrial Intelligence and Systems Optimization, Northeastern University, Shenyang 110819, China}
\affiliation{Key Laboratory of Data Analytics and Optimization for Smart Industry (Northeastern University), Ministry of Education, Shenyang 110819, China}

\begin{abstract}
Multi-messenger gravitational-wave (GW) observation for binary neutron star merger events could provide a rather useful tool to explore the evolution of the universe. In particular, for the third-generation GW detectors, i.e., the Einstein Telescope (ET) and the Cosmic Explorer (CE), proposed to be built in Europe and the U.S., respectively, lots of GW standard sirens with known redshifts could be obtained, which would exert great impacts on the cosmological parameter estimation. The total neutrino mass could be measured by cosmological observations, but such a measurement is model-dependent and currently only gives an upper limit. In this work, we wish to investigate whether the GW standard sirens observed by ET and CE could help improve the constraint on the neutrino mass, in particular in the interacting dark energy (IDE) models. We find that the GW standard siren observations from ET and CE can only slightly improve the constraint on the neutrino mass in the IDE models, compared to the current limit. The improvements in the IDE models are weaker than those in the standard cosmological model. Although the limit on neutrino mass can only be slightly updated, the constraints on other cosmological parameters can be significantly improved by using the GW observations.

\end{abstract}
\maketitle

\section{Introduction}\label{sec1}

In the recent two decades, the study of cosmology has entered the era of precision cosmology. A standard model of cosmology has been established, usually called the $\Lambda$ cold dark matter ($\Lambda$CDM) model. The measurements of cosmic microwave background (CMB) anisotropies from the Planck satellite mission have constrained the six primary parameters of the $\Lambda$CDM model with unprecedented precision. However, with the measurement precisions of the cosmological parameters improved, some puzzling issues appeared. For example, the inferred values of the Hubble constant from the Planck observation of the CMB anisotropies (based on the $\Lambda$CDM model) \cite{Planck:2018vyg} and from the Cepheid-supernova distance ladder measurement \cite{Riess:2019cxk} are inconsistent, with the tension between them more than 4$\sigma$ significance \cite{Riess:2019cxk}.
Namely, there is an inconsistency of measurements between the early and late universe, which is the so-called ``Hubble tension'' problem. The Hubble tension recently has been widely discussed in the literature (see, e.g., Refs.~\cite{Riess:2019qba,Verde:2019ivm,Li:2013dha,Zhang:2014dxk,Gao:2021xnk,cai:2020,Cai:2021wgv,Zhao:2017urm,Guo:2018ans,Guo:2019dui,Yang:2018euj,Yan:2019gbw,Vagnozzi:2019ezj,DiValentino:2019jae,DiValentino:2019ffd,Liu:2019awo,Zhang:2019cww,Ding:2019mmw,Feng:2019jqa,Lin:2020jcb,Xu:2020uws,Li:2020tds,Hryczuk:2020jhi,Wang:2021kxc,Vagnozzi:2021tjv,Vagnozzi:2021gjh,DiValentino:2021izs,Dainotti:2021pqg,Ren:2022aeo}).
Furthermore, theoretically, for the $\Lambda$CDM model, the cosmological constant $\Lambda$, which is equivalent to the density of vacuum energy, has always been suffering from serious theoretical challenges, such as the ``fine-tuning'' and ``cosmic coincidence'' problems \cite{Sahni:1999gb,Bean:2005ru}. Thus, it is hard to say that the $\Lambda$CDM model with only six base parameters is the eventual model of cosmology. All of these facts actually imply that the $\Lambda$CDM model needs to be further extended and some extra parameters concerning new physics need to be introduced into the new models. Of course, some novel cosmological probes should also be further developed.

To extend the $\Lambda$CDM cosmology, the primary idea is to consider the dynamical dark energy with the dark-energy density no longer a constant. In this class of models,
the simplest one is the model with a dark energy having a constant equation-of-state (EoS) parameter, $w=p_{\rm de} / \rho_{\rm de}=$ constant, which is usually called the $w$CDM model.
For some popular dark energy models, see, e.g., Refs.~\cite{Chevallier:2000qy,Linder:2002et,Huterer:2000mj,Wetterich:2004pv,Jassal:2004ej,Upadhye:2004hh,Xia:2004rw,Zhang:2005yz,Zhang:2005hs,Zhang:2006qu,Linder:2006sv,Zhang:2007sh,Lazkoz:2010gz,Ma:2011nc,Li:2011dr,Li:2012vn,Li:2012via,Cai:2015emx}. There is also a class of models known as the interacting dark energy (IDE) models in which some direct, non-gravitational interaction between dark energy and dark matter is considered.
The interaction between dark sectors could help resolve (or alleviate) the coincidence problem of dark energy, and also can help alleviate the Hubble tension. The IDE models have been widely studied and deeply explored till now (see, e.g., Refs.~\cite{Amendola:1999er,Amendola:1999qq,Tocchini-Valentini:2001wmi,Amendola:2001rc,Comelli:2003cv,Chimento:2003iea,Cai:2004dk,Zhang:2004gc,Ferrer:2004nv,Zhang:2005rj,Sadjadi:2006qp,Barrow:2006hia,Sasaki:2006kq,Abdalla:2007rd,Bean:2007ny,Guo:2007zk,Zhang:2009un,Caldera-Cabral:2009hoy,Valiviita:2009nu,He:2009mz,Koyama:2009gd,Li:2009zs,Xia:2009zzb,Cai:2009ht,Li:2011ga,Zhang:2012uu,Xu:2013jma,Zhang:2013zyn,Wang:2013qy,Salvatelli:2013wra,Yang:2014okp,Wang:2014oga,Faraoni:2014vra,Fan:2015rha,Yang:2015tzc,Duniya:2015nva,Murgia:2016ccp,Costa:2016tpb,Xia:2016vnp,Guo:2017hea,Zhang:2017ize,Feng:2018yew,Yang:2018euj,Guo:2018ans,Zhao:2018fjj,Feng:2019mym,Li:2019loh,Li:2020gtk}).

Currently, the mainstream cosmological probes mainly include, e.g., the CMB anisotropies, baryon acoustic oscillations (BAO), type Ia supernovae (SN), direct determination of the Hubble constant ($H_{0}$), weak gravitational lensing, redshift space distortions, and clusters of galaxies. The combinations of these cosmological data based on the electromagnetic (EM) observations have provided precise measurements for the base cosmological parameters. But for some extended parameters beyond the standard $\Lambda$CDM model, e.g., the EoS parameter of dark energy, the tensor-to-scalar ratio, and the total mass of neutrinos, they still cannot be precisely measured. Therefore, on one hand, the EM observations should be further greatly developed, and on the other hand, some novel cosmological probes are also needed to be developed in the future. In the next few decades, the gravitational-wave (GW) standard siren observation is one of the most promising cosmological probes.

The detection of the GW event GW170817 \cite{LIGOScientific:2017vwq} from the binary neutron star (BNS) merger initiated the multi-messenger astronomy era. Because in this event, not only GWs but also the EM signals in various bands were detected for the same transient source \cite{LIGOScientific:2017zic,LIGOScientific:2017ync}. From the analysis of the waveform of GW, one can obtain the absolute luminosity distance to the source. Furthermore, the redshift of the source can also be determined by identifying the EM counterpart of the GW source. With the known distance and redshift of a celestial source, a distance--redshift relation can be established, which can be used to explore the expansion history of the universe \cite{Schutz:1986gp}. Such a tool of exploring the evolution of the universe provided by GWs is called ``standard sirens'' (note here that the case having EM counterparts is usually referred to as bright sirens, to be differentiated with the case of dark sirens without EM counterparts) \cite{Holz:2005df}.

The main advantage of the GW standard siren method is that the absolute luminosity distances can be measured. This is obviously superior to the SN observation, in that the latter can only measure the ratio of luminosity distances at different redshifts. In addition, the GW observation can observe much higher redshift events, compared to the SN observation.

It is indisputable that the GW standard siren will be developed into a powerful cosmological probe in the future. The third-generation (3G) ground-based GW detectors, such as the Einstein Telescope (ET) \cite{ET-web,Punturo:2010zz} in Europe and the Cosmic Explorer (CE) \cite{CE-web,Evans:2016mbw} in the United States, have been proposed. In the 2030s, ET will be brought into operation. CE will start its observation in the mid-2030s. The 3G ground-based GW detectors have a much wider detection-frequency range and a much better detection sensitivity, which can observe much more BNS events at much deeper redshifts. Recently, the GW standard sirens have been widely discussed in the literature \cite{Cai:2016sby,Cai:2017aea,Cai:2017plb,Zhang:2019ylr,Zhao:2018gwk,Du:2018tia,Cai:2018rzd,Yang:2019bpr,Yang:2019vni,Bachega:2019fki,Chang:2019xcb,He:2019dhl,Liu:2017xef,Berti:2018cxi,Liu:2018sia,Will:1994fb,Zhao:2019gyk,Wang:2021srv,Qi:2021iic,Jin:2021pcv,Buchmuller:2019gfy,Borhanian:2022czq,Colgain:2022xcz,Zhu:2021bpp,deSouza:2021xtg,Wang:2022oou,Jin:2022qnj,Wu:2022dgy} {(see Ref.~\cite{Bian:2021ini} for a recent review)}. It is found that the GW standard siren observations from ET and CE would play an important role in the cosmological parameter estimation \cite{Zhang:2019ple,Zhang:2018byx,Wang:2018lun,Zhang:2019loq,Li:2019ajo,Jin:2020hmc}.

In cosmology, neutrinos play a crucial role in helping shape the large-scale structure and the expansion history of the universe.
The phenomenon of neutrino oscillation indicates that neutrinos have masses and there are mass splittings between different neutrino species. However, it is extremely difficult to measure the absolute masses of neutrinos. Neutrino oscillation experiments cannot measure the absolute neutrino masses, but can only give the squared
mass differences between the different mass eigenstates of neutrinos. The solar and reactor experiments give $\Delta m_{21}^{2} \simeq 7.5 \times 10^{-5}$ $\rm eV^{2}$ and the atmospheric and accelerator beam experiments give $|\Delta m_{31}^{2}| \simeq 2.5 \times 10^{-3}$ $\rm eV^{2}$ \cite{ParticleDataGroup:2014cgo,Xing:2020ijf}. Thus, there are two possible mass hierarchies of the neutrino mass spectrum, namely, the normal hierarchy (NH) with $m_{1} < m_{2} \ll m_{3}$ and the inverted hierarchy (IH) with $m_{3} \ll m_{1} < m_{2}$.
In addition, in some cases one also considers the cosmological models of neglecting the neutrino mass splittings, namely $m_1=m_2=m_3$, which is usually called the degenerate hierarchy (DH).

Although the neutrino masses can hardly be measured by particle physics experiments, they can be effectively constrained by the cosmological observations. This is because massive neutrinos can exert some impacts on the evolution of the universe. Using the current cosmological observations, an upper limit on the total neutrino mass $\sum m_\nu$ can be obtained. So far, the most stringent limit on the total neutrino mass comes from the Planck 2018 CMB observation, and the combination CMB+BAO+SN gives the $95\%$ CL upper limit $\sum m_\nu<0.12$ eV, for the DH case in the $\Lambda$CDM model. See e.g. Refs.~\cite{Hu:1997mj,Reid:2009nq,Thomas:2009ae,Carbone:2010ik,Huo:2011ve,Wang:2012vh,Li:2012spm,Audren:2012vy,Riemer-Sorensen:2013jsa,Font-Ribera:2013rwa,Zhang:2014nta,Zhang:2014ifa,Zhou:2014fva,Planck:2015fie,Zhang:2015rha,Geng:2015haa,Lu:2016hsd,Kumar:2016zpg,Xu:2016ddc,Vagnozzi:2017ovm,Zhang:2017rbg,Zhao:2017jma,Vagnozzi:2018jhn,Li:2017iur,Wang:2017htc,Feng:2017usu,Zhao:2016ecj,Zhang:2015uhk,Huang:2015wrx, Wang:2016tsz,Giusarma:2016phn,Allahverdi:2016fvl,Gariazzo:2018pei,RoyChoudhury:2018gay,Han:2017wnk,Zhang:2019ipd,DiazRivero:2019ukx,Zhang:2020mox} for studies on neutrino mass in cosmology.


In a recent forecast \cite{Wang:2018lun}, it was shown that the standard sirens observed by the ET can be used to improve the constraints on the total neutrino mass in the $\Lambda$CDM model. Using 1000 GW standard siren data points of the BNS merger events, it is found that the upper limits on $\sum m_\nu$ can be tightened by about $10\%$ \cite{Wang:2018lun}. However, weighing neutrinos in cosmology depends on the cosmological model considered, and thus one would be curious about whether the role the GW data play in helping measure the neutrino mass will change if an extension to the $\Lambda$CDM model is considered. In this work, we consider the IDE models, and we wish to see what will happen on measuring neutrino mass when the IDE models are considered.

In an IDE model, the energy conservation equations for dark energy and CDM satisfy
\begin{align}
&\dot{\rho}_{\mathrm{de}}=-3 H(1+w) \rho_{\mathrm{de}}+Q, \\
&\dot{\rho}_{\mathrm{c}}=-3 H \rho_{\mathrm{c}}-Q,
\end{align}
where $Q$ is the energy transfer rate, $\rho_{\rm de}$ and $\rho_c$ represent the energy densities of dark energy and CDM, respectively, $H$ is the Hubble parameter, and a dot represents the derivative with respect to the cosmic time $t$. In this work, we consider the interaction form of $Q=\beta H \rho_{\mathrm{c}}$, where $\beta$ is a dimensionless coupling parameter. Here, $\beta>0$ and $\beta<0$ means CDM decaying into dark energy and dark energy decaying into CDM, respectively.

In this work, we consider the IDE versions of the $\Lambda$CDM and $w$CDM models, which are called the I$\Lambda$CDM and I$w$CDM models. We will discuss the cosmological parameter estimation in the I$\Lambda$CDM+$\sum m_{\nu}$ and I$w$CDM+$\sum m_{\nu}$ models. Moreover, we will consider the three neutrino mass hierarchy cases, i.e., the NH, IH, DH cases. To avoid the perturbation divergence problem in the IDE models, in this work we employ the extended parameterized post-Friedmann (ePPF) framework \cite{Li:2014eha,Li:2014cee} to calculate the perturbations of dark energy.

We simulate the GW standard siren data observed by ET and CE, and we use these simulated GW data to investigate how well they can be used to improve the constraints on the neutrino mass as well as other cosmological parameters on the basis of the current CMB+BAO+SN constraints.



{The rest of this paper is organized as follows. In Sec.~\ref{sec2.1}, we introduce the methods of simulating the GW standard siren data. In Sec.~\ref{sec2.2} we describe the EM cosmological observations used in this work. In Sec.~\ref{sec2.3}, we briefly describe the methods of constraining cosmological parameters. In Sec.~\ref{sec3}, we give the constraint results and make some relevant discussions. The conclusion is given in Sec.~\ref{sec4}.}

\section{Method and data}\label{sec2}

In this section, we first introduce the method of simulating the GW standard siren data from ET and CE. Then, we describe the current mainstream EM cosmological observations used in this work. Finally, we briefly introduce the method of constraining cosmological parameters.

\subsection{Simulation of the GW standard sirens}\label{sec2.1}


{The primary GW sources in the detection frequency band of the ground-based GW detectors are the mergers of BNS, binary stellar-mass black hole (BBH), and so on. The BNS mergers could produce rich EM signals \cite{Li:1998bw} that can be detected by the EM observatories, thus enabling precise redshift measurements. Owing to the fact that there are no EM signals produced in the process of the BBH mergers, their redshifts could not be precisely measured through the detection of the EM counterparts. Hence, in this work, we only simulate the GW standard sirens from the BNS mergers.}

\begin{figure}[!htp]
\includegraphics[width=8.6cm]{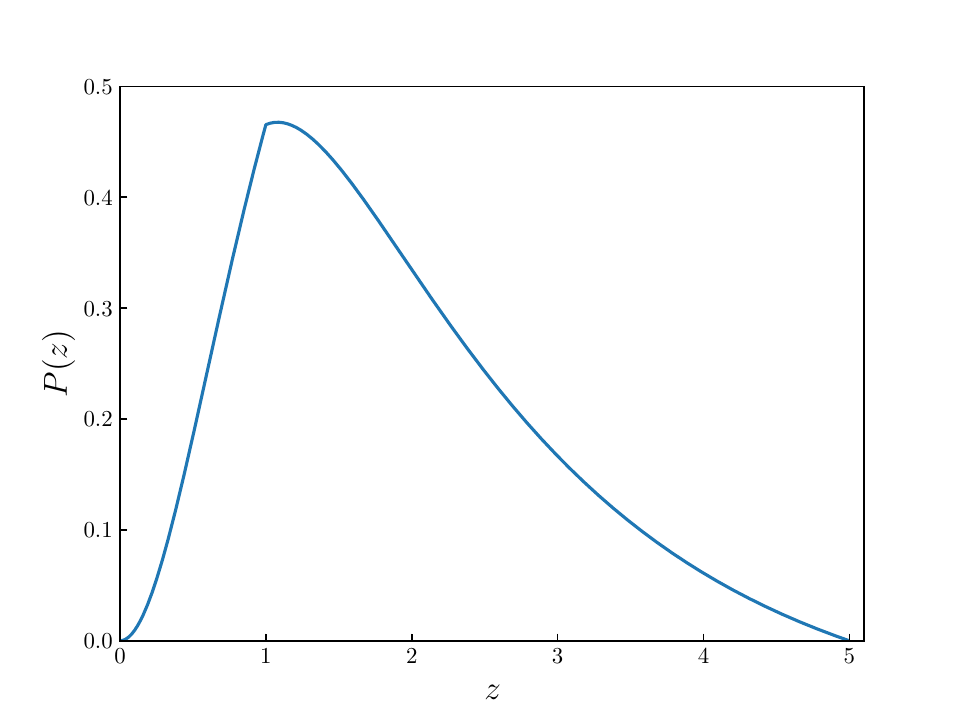}\ \hspace{1cm}
\centering \caption{\label{fig1} The redshift distribution of BNS mergers.}
\end{figure}

Following Refs.~\cite{Zhao:2010sz,Cai:2016sby}, the redshift distribution of the BNS mergers takes the form
\begin{equation}
P(z)\propto \frac{4\pi d_{\rm C}^2(z)R(z)}{H(z)(1+z)},
\label{equa:pz}
\end{equation}
where $d_{\rm C}(z)$ is the comoving distance at the redshift $z$, and $R(z)$ represents the redshift evolution of the burst rate, which takes the form \cite{Schneider:2000sg,Cutler:2009qv,Cai:2016sby}
\begin{equation}
R(z)=\begin{cases}
1+2z, & z\leq 1, \\
\frac{3}{4}(5-z), & 1<z<5, \\
0, & z\geq 5.
\end{cases}
\label{equa:rz}
\end{equation}
In Fig.~\ref{fig1}, we show the redshift distribution of BNS mergers.

Considering the transverse-traceless gauge, the strain $h(t)$ in the GW interferometers can be described by two independent polarization amplitudes, $h_+(t)$ and $h_\times(t)$,
\begin{equation}
h(t)=F_+(\theta, \phi, \psi)h_+(t)+F_\times(\theta, \phi, \psi)h_\times(t),
\end{equation}
where $F_{+}$ and $F_{\times}$ are the antenna response functions, $\theta$ and $\phi$ describe the location of the GW source relative to the GW detector, and $\psi$ is the polarization angle.

The antenna response functions of ET are \cite{Zhao:2010sz}
 \begin{align}
F_+^{(1)}(\theta, \phi, \psi)=&~~\frac{{\sqrt 3 }}{2}\bigg[\frac{1}{2}\big(1 + {\cos ^2}(\theta )\big)\cos (2\phi )\cos (2\psi ) \nonumber\\
                              &~~- \cos (\theta )\sin (2\phi )\sin (2\psi )\bigg],\nonumber\\
F_\times^{(1)}(\theta, \phi, \psi)=&~~\frac{{\sqrt 3 }}{2}\bigg[\frac{1}{2}\big(1 + {\cos ^2}(\theta )\big)\cos (2\phi )\sin (2\psi ) \nonumber\\
                              &~~+ \cos (\theta )\sin (2\phi )\cos (2\psi )\bigg].
\label{equa:F1}
\end{align}
Since ET has three interferometers with $60^\circ$ inclined angles between each other, the other two response functions are $F_{+,\times}^{(2)}(\theta, \phi, \psi)=F_{+,\times}^{(1)}(\theta, \phi+2\pi/3, \psi)$ and $F_{+,\times}^{(3)}(\theta, \phi, \psi)=F_{+,\times}^{(1)}(\theta, \phi+4\pi/3, \psi)$.

For CE, the antenna response functions are
\begin{align}
F_+(\theta, \phi, \psi)=&~~\frac{1}{2}\big(1 + {\cos ^2}(\theta )\big)\cos (2\phi )\cos (2\psi ) \nonumber\\
                              &~~- \cos (\theta )\sin (2\phi )\sin (2\psi ),\nonumber\\
F_\times(\theta, \phi, \psi)=&~~\frac{1}{2}\big(1 + {\cos ^2}(\theta )\big)\cos (2\phi )\sin (2\psi ) \nonumber\\
                              &~~+ \cos (\theta )\sin (2\phi )\cos (2\psi ).
\label{equa:F2}
\end{align}

{We consider the waveform in the inspiralling stage for a non-spinning BNS system. Here we adopt the restricted post-Newtonian (PN) approximation and calculate the waveform to the 3.5 PN order \cite{Sathyaprakash:2009xs,Blanchet:2004bb},}
\begin{equation}
\tilde{h}(f)=\mathcal{A}f^{-7/6}\exp[{\rm i}(2\pi ft_{\rm c}-\pi/4-2\psi_{\rm c}+2{\Psi}(f/2)-\varphi_{(2.0)})],
\label{equa:hf}
\end{equation}
where the Fourier amplitude $\mathcal{A}$ is given by
\begin{align}
\mathcal{A}=&~~\frac{1}{d_{\rm L}}\sqrt{F_+^2\big(1+\cos^2(\iota)\big)^2+4F_\times^2\cos^2(\iota)}\nonumber\\
            &~~\times \sqrt{5\pi/96}\pi^{-7/6}\mathcal{M}_{\rm c}^{5/6},
\label{equa:A}
\end{align}
where $\mathcal{M}_{\rm c}=(1+z)M \eta^{3/5}$ is the observed chirp mass, $M=m_1+m_2$ is the total mass of the coalescing binary system with $m_1$ and $m_2$ being the component masses, $\eta=m_1 m_2/M^2$ is the symmetric mass ratio, $d_{\rm L}$ is the luminosity distance to the GW source, $\iota$ is the inclination angle between the binary's orbital angular momentum and the line of sight, $t_{\rm c}$ is the coalescence time, and $\psi_{\rm c}$ is the coalescence phase. The definitions of the functions $\Psi$ and $\varphi_{(2.0)}$ can refer to Refs.~\cite{Sathyaprakash:2009xs,Blanchet:2004bb}.

The signal-to-noise ratio (SNR) for the network of $N$ (i.e., $N$ = 3 for ET and $N$ = 1 for CE) independent interferometers can be calculated by
\begin{equation}
\rho=\sqrt{\sum\limits_{i=1}^{N}(\rho^{(i)})^2},
\label{euqa:rho}
\end{equation}
where $\rho^{(i)}=\sqrt{\langle \tilde{h}^{(i)},\tilde{h}^{(i)}\rangle}$. The inner product is defined as
\begin{equation}
\left\langle{a,b}\right\rangle=4\int_{f_{\rm lower}}^{f_{\rm upper}}\frac{ a(f) b^\ast(f)+ a^\ast(f) b(f)}{2}\frac{{\rm d}f}{S_{\rm n}(f)},
\label{euqa:product}
\end{equation}
where $f_{\rm lower}$ is the lower cutoff frequency ($f_{\rm lower}=1$ Hz for ET and $f_{\rm lower}=5$ Hz for CE), $f_{\rm upper}=2/(6^{3/2}2\pi M_{\rm obs})$ is the frequency at the last stable orbit
with $M_{\rm obs}=(m_1+m_2)(1+z)$ \cite{Zhao:2010sz}, and $S_{\rm n}(f)$ is the one-side noise power spectral density (PSD). We adopt PSD of ET from Ref.~\cite{ETcurve-web} and that of CE from Ref.~\cite{CEcurve-web}. In this work, we choose the threshold of SNR to be 8 in our simulation.

{For the 3G ground-based GW detectors, a few $\times 10^5$ BNS merger events per year could be detected, but only about 0.1\% of them may have $\gamma$-ray bursts toward us \cite{Yu:2021nvx}. Recently, in Ref.~\cite{Chen:2020zoq} Chen \emph{et al.} made a forecast and showed that 910 GW standard siren events could be observed by a 10-year observation of CE and Swift++. Therefore, in our forecast in the present work, for ET and CE, we simulate 1000 GW standard sirens from BNS mergers based on the 10-year observation. }

We consider three measurement errors of $d_{\rm L}$, consisting of the instrumental error $\sigma_{d_{\rm L}}^{\rm inst}$, the weak-lensing error $\sigma_{d_{\rm L}}^{\rm lens}$, and the peculiar velocity error $\sigma_{d_{\rm L}}^{\rm pv}$. Therefore, the total error of $d_{\rm L}$ is
\begin{align}
\sigma_{d_{\rm L}}&~~=\sqrt{(\sigma_{d_{\rm L}}^{\rm inst})^2+(\sigma_{d_{\rm L}}^{\rm lens})^2+(\sigma_{d_{\rm L}}^{\rm pv})^2}.\label{total}
\end{align}


We first use the Fisher information matrix to calculate $\sigma_{d_{\rm L}}^{\rm inst}$. {For a GW event, when using Fisher information matrix to estimate $\sigma_{d_{\rm L}}^{\rm inst}$, we consider a $9\times 9$ Fisher information matrix consisting of nine parameters of a GW source ($d_{\rm L}$, $\mathcal{M}_{\rm c}$, $\eta$, $\theta$, $\phi$, $\iota$, $t_{\rm c}$, $\psi_{\rm c}$, $\psi$), and thus the correlations between the nine parameters are considered in the analysis.}
For a network of $N$ independent interferometers, the Fisher information matrix can be written as
\begin{equation}
\boldsymbol{F}_{ij}=\left\langle\frac{\partial \boldsymbol{h}(f)}{\partial \theta_i},  \frac{\partial \boldsymbol{h}(f)}{\partial \theta_j}\right\rangle,
\end{equation}
with $\boldsymbol{h}$ given by
\begin{equation}
\boldsymbol{h}(f)=\left[\tilde{h}_1 (f),\tilde{h}_2 (f),\cdots,\tilde{h}_N (f)\right],
\end{equation}
where $\theta_i$ denotes nine parameters ($d_{\rm L}$, $\mathcal{M}_{\rm c}$, $\eta$, $\theta$, $\phi$, $\iota$, $t_{\rm c}$, $\psi_{\rm c}$, $\psi$) for a GW event. Then we have
\begin{equation}
\Delta \theta_i =\sqrt{(F^{-1})_{ii}},
\end{equation}
where $F_{ij}$ is the total Fisher information matrix for the network of $N$ interferometers. Note that here $\sigma_{d_{\rm L}}^{\rm inst}=\Delta \theta_1$.

{The error caused by weak lensing is adopted from Refs.~\cite{Hirata:2010ba,Speri:2020hwc},
\begin{equation}
\sigma_{d_{\rm L}}^{\rm lens}(z)=F_{\rm {delens}}(z)\times d_{\rm L}(z)\times 0.066\bigg[\frac{1-(1+z)^{-0.25}}{0.25}\bigg]^{1.8}.\label{lens}
\end{equation}
Here, we consider a delensing factor $F_{\rm {delens}}$. We use dedicated matter surveys along the line of sight of the GW event in order to estimate the lensing magnification distribution, which can remove part of the uncertainty due to weak lensing. This reduces the weak lensing uncertainty. The delensing factor is given by
\begin{equation}
F_{\rm {delens}}(z)=1-\frac{0.3}{\pi / 2} \arctan \left(z / 0.073\right).
\end{equation}}

{The error caused by the peculiar velocity of the GW source is given by \cite{Kocsis:2005vv}
\begin{equation}
\sigma_{d_{\rm L}}^{\rm pv}(z)=d_{\rm L}(z)\times \bigg[ 1+ \frac{c(1+z)^2}{H(z)d_{\rm L}(z)}\bigg]\frac{\sqrt{\langle v^2\rangle}}{c},\label{pv}
\end{equation}
where $H(z)$ is the Hubble parameter. $\sqrt{\langle v^2\rangle}$ is the peculiar velocity of the GW source and we roughly set $\sqrt{\langle v^2\rangle}=500\ {\rm km\ s^{-1}}$.}

For each simulated GW source, {the sky location, the binary inclination, the coalescence phase, and the polarization angle are randomly chosen in the ranges of ${\rm cos}\,\theta\in [-1,1]$, $\phi\in [0,360^\circ]$, $\iota\in [0,20^\circ]$, $\psi_{\rm c}\in [0,360^\circ]$, and $\psi\in [0,360^\circ]$. The mass of an NS is randomly chosen in the range of $[1, 2]$ $M_\odot$. Without loss of generality, the merger time is chosen to $t_{\rm c}=0$ in our analysis. Here we wish to note that the inclination angle should be randomly chosen in the range of ${\rm cos}\,\iota\in [-1,1]$ when simulating isotropic GW sources. However, in this work, we simulate GW events by detecting short $\gamma$-ray bursts (SGRBs) to determine sources' redshifts. Owing to the fact that SGRBs are strongly beamed \cite{Rezzolla:2011da}, the detectable inclination angle is about $\iota\leq 20^\circ$ \cite{li2015extracting,Yu:2021nvx}. Hence, in the present work, we set the inclination angle to be in the range of $[0,20^\circ]$. This is an ideal treatment, but for this work, since the number of simulated GW standard sirens is fixed, it has little effect on showing the impact of GW standard sirens on breaking cosmological parameter degeneracies and improving constraints on the cosmological parameters.}


\begin{figure}[!htp]
\includegraphics[width=0.9\linewidth,angle=0]{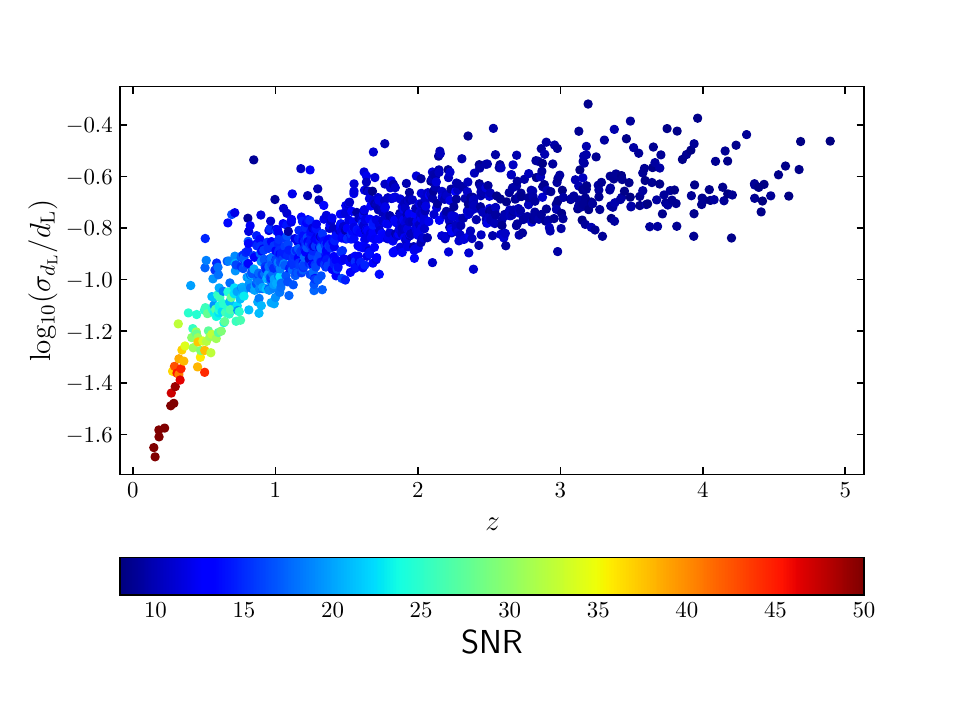}\ \hspace{1cm}
\includegraphics[width=0.9\linewidth,angle=0]{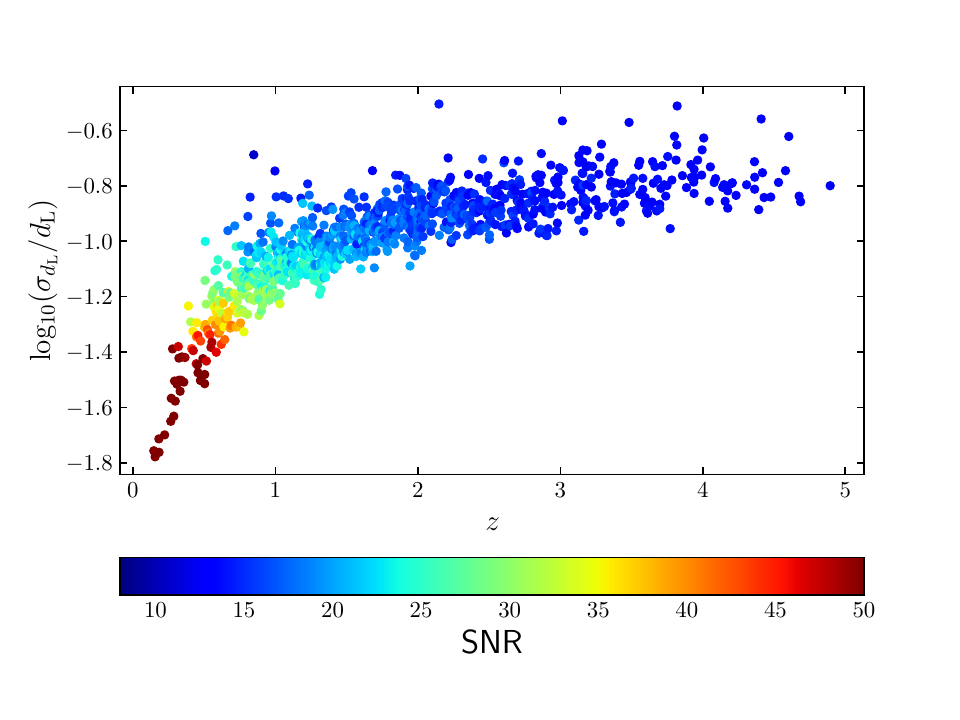}\ \hspace{1cm}
\centering \caption{\label{fig2} {Distribution of $\sigma_{d_{\rm L}}/d_{\rm L}$ as a function of redshift. The color indicates SNR of the simulated GW standard sirens. {\it Upper}: 1000 GW standard sirens from a 10-year observation of ET. {\it Lower}: 1000 GW standard sirens from a 10-year observation of CE.}}
\end{figure}

In Fig.~\ref{fig2}, we show the $\sigma_{d_{\rm L}}/d_{\rm L}$ scatter plot of the simulated standard siren data from ET (upper panel) and CE (lower panel).\footnote{Here we wish to note that in the colorbars of Fig.~2, we set the maximum value to be 50, i.e., a GW event with SNR greater than 50 has the same color as the GW event with SNR of 50. In fact, many red dots have SNRs greater than 50, e.g., SNR at $z=0.146$ for ET is 105.9, while for CE is 158.2. The reason of the design is that we can better highlight the comparison between ET and CE from the figure. If we set the maximal value of SNR higher (to about 100), almost all the data points are in blue, so it is difficult to visually compare SNRs of the two detectors from the figure.} We can see that: (i) the total errors of $d_{\rm L}$ from CE are smaller than those from ET; (ii) SNR of CE is higher than that of ET at the same redshift.

\begin{figure*}[!htp]
\includegraphics[scale=0.5]{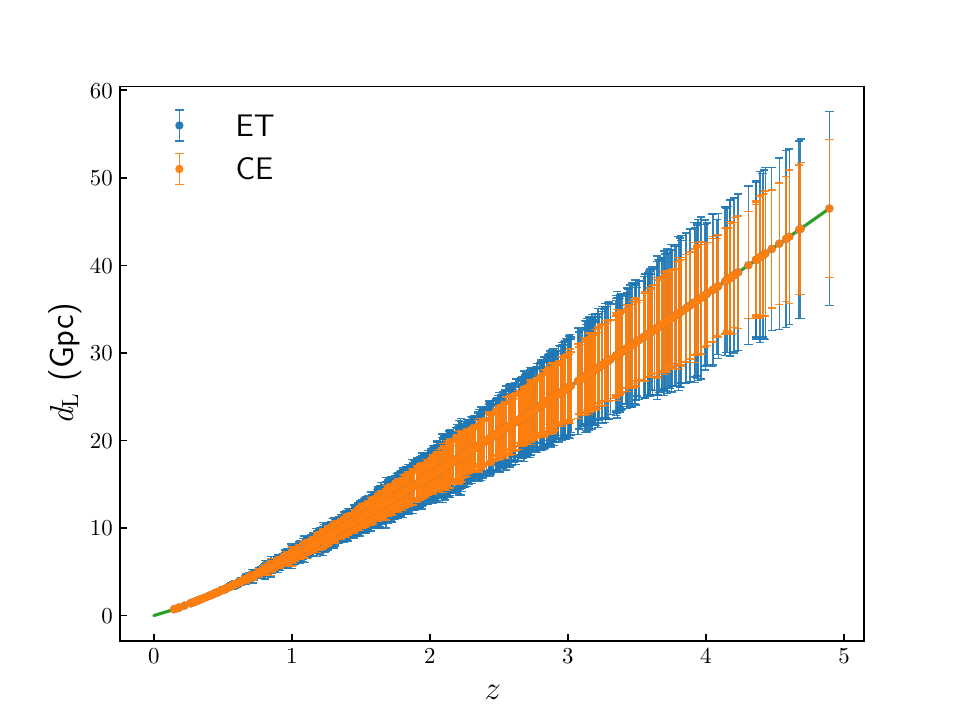}\ \hspace{1cm}
\includegraphics[scale=0.5]{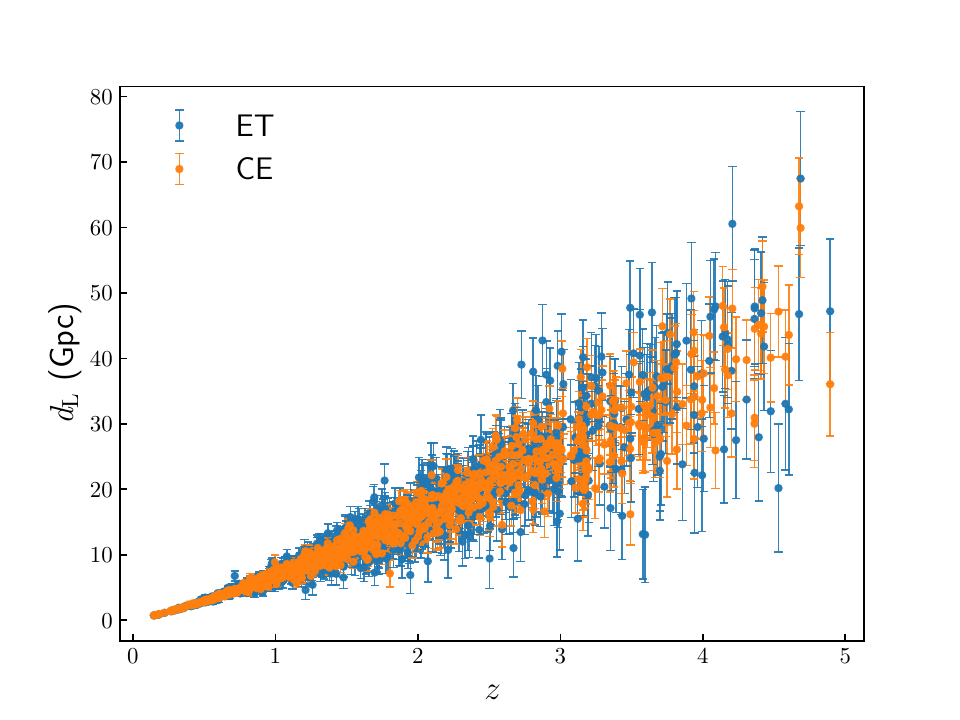}\ \hspace{1cm}
\centering \caption{\label{fig3} {The simulated GW standard siren data points observed by ET and CE. The blue data points represent the 1000 standard sirens from the 10-year observation of ET and the orange data points represent the 1000 standard sirens from the 10-year observation of CE. {\it Left}: the standard siren data points without Gaussian randomness, where the central values of the luminosity distances are calculated by the fiducial cosmological model, and the solid green line represents the $d_{\rm L}(z)$ curve predicted by the fiducial model. {\it Right}: the standard siren data points with Gaussian randomization, reflecting the fluctuations in measured values resulted from actual observations.}}
\end{figure*}

{In Fig.~\ref{fig3}, we show the simulated GW standard sirens from ET and CE. In the left panel, we show the standard siren data points without Gaussian randomness, where the central value of the luminosity distance is calculated by the fiducial cosmological model. In the right panel, in order to reflect the fluctuations in measured values resulting from actual observations, we show the standard siren data points with Gaussian randomization ({the central values are populated according to a Gaussian distribution with mean being $d_{\rm L}$ and standard deviation being $\sigma_{d_{\rm L}}$}). {In principle, the right panel is more representative of actual observational data, but the central values of $d_{\rm L}$ have no effect on determining the absolute errors of cosmological parameters. Therefore, we use the data points in the left panel to constrain the cosmological models, because this is more helpful in investigating how the parameter degeneracies are broken to improve measurement precisions of cosmological parameters.} We can clearly see that the measurement errors of $d_{\rm L}$ from CE are smaller than those from ET, because CE has a better sensitivity than ET.}

\subsection{Other cosmological observations}\label{sec2.2}

In this work, we consider three current mainstream EM cosmological observations, including CMB, BAO, and SN.
For the CMB data, we consider the Planck TT, TE, EE spectra at $\ell \geq 30$, the low-$\ell$ temperature Commander likelihood, and the low-$\ell$ SimAll EE likelihood from the Planck 2018 release \cite{Planck:2018vyg}. For the BAO data, we consider the measurements from 6dFGS ($z_{\rm eff}=0.106$) \cite{Beutler:2011hx}, SDSS-MGS ($z_{\rm eff}=0.15$) \cite{Ross:2014qpa}, and BOSS DR12 ($z_{\rm eff}=0.38$, 0.51, and 0.61) \cite{BOSS:2016wmc}. For the SN data, we use the latest Pantheon sample, which is comprised of 1048 data points from the Pantheon compilation \cite{Scolnic:2017caz}.

\subsection{Method of constraining cosmological parameters}\label{sec2.3}

To resolve the large-scale instability problem in the IDE cosmology \cite{Valiviita:2008iv}, we apply the ePPF approach \cite{Li:2014eha,Li:2014cee} for the IDE scenario so that the whole parameter space of IDE models can be explored without any divergence of the dark-energy perturbation. In this work, we employ the modified version of the available Markov-Chain Monte Carlo package {\tt CosmoMC} \cite{Lewis:2002ah}, with the ePPF code \cite{Li:2014eha,Li:2014cee} inserted, to constrain the neutrino mass and other cosmological parameters. In order to show the impacts of GW data from CE and ET on constraining cosmological parameters, we use CMB+BAO+SN, CMB+BAO+SN+ET, and CMB+BAO+SN+CE to make our analysis. For convenience, we use CBS to standard for CMB+BAO+SN in the following.


For the GW standard siren observation with $N$ data points, the $\chi^2$ function can be written as
\begin{align}
\chi_{\rm GW}^2=\sum\limits_{i=1}^{N}\left[\frac{{d}_L^i-d_{\rm L}({z}_i;\vec{\Omega})}{{\sigma}_{d_{\rm L}}^i}\right]^2,
\label{equa:chi2}
\end{align}
where ${z}_i$, ${d}_L^i$, and ${\sigma}_{d_{\rm L}}^i$ are the $i$th GW event's redshift, luminosity distance, and the measurement error of the luminosity distance, respectively. $\vec{\Omega}$ represents the set of cosmological parameters.

When considering the combination of the current EM observations and the GW standard siren observations, the total $\chi^{2}_{\rm tot}$ function is
\begin{equation}
\chi^{2}_{\rm tot}=\chi^{2}_{\rm CMB}+\chi^{2}_{\rm BAO}+\chi^{2}_{\rm SN}+\chi^{2}_{\rm GW}.
\end{equation}

\begin{figure*}[!htp]
\includegraphics[width=0.45\textwidth]{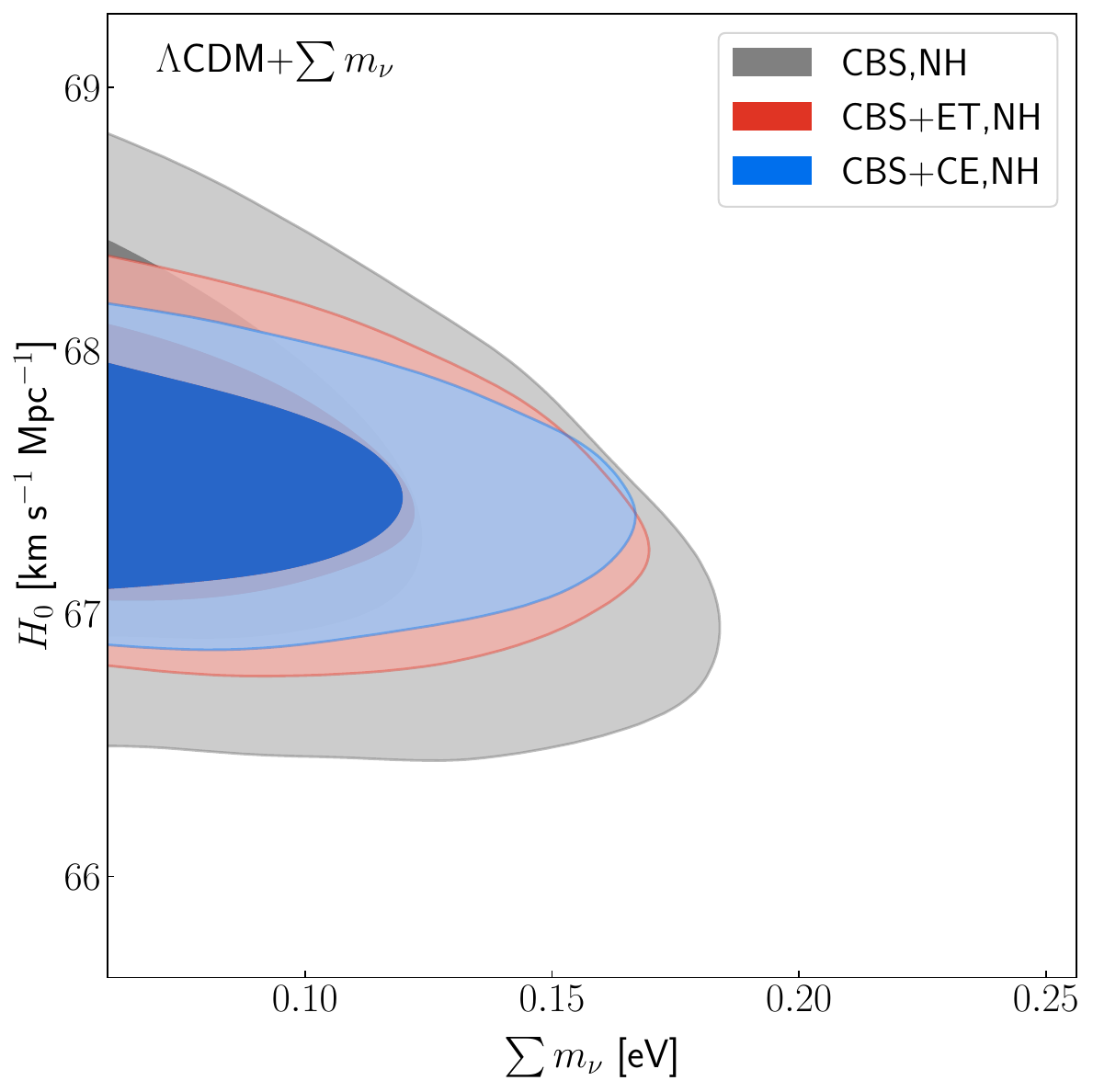} \ \hspace{1cm}
\includegraphics[width=0.45\textwidth]{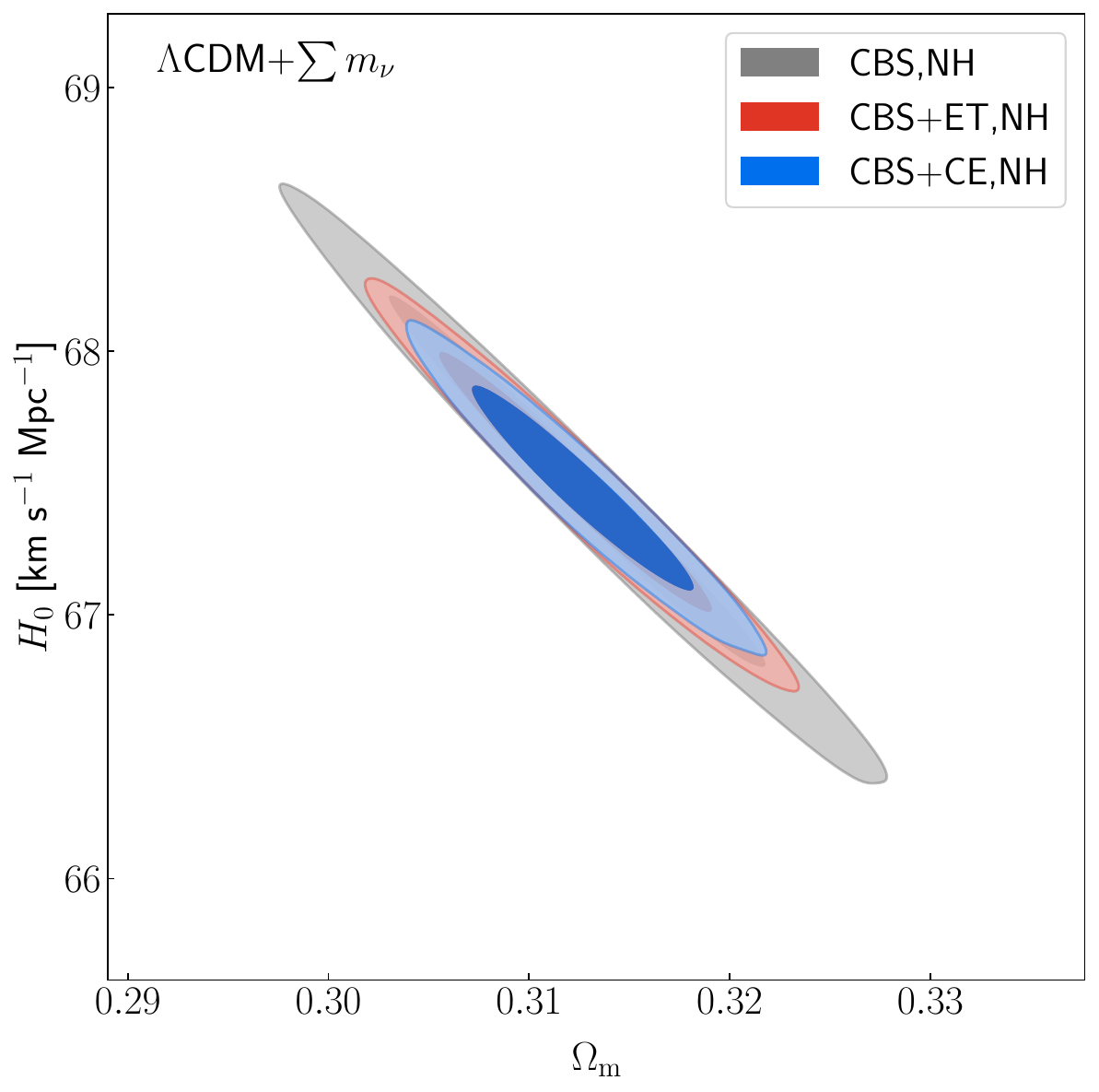}\ \hspace{1cm}
\centering \caption{\label{fig4} {Two-dimensional marginalized contours (68.3\% and 95.4\% confidence level) in the $\sum m_\nu$--$H_0$ and $\Omega_{\rm m}$--$H_0$ planes using the CBS, CBS+ET, and CBS+CE data. Here CBS stands for CMB+BAO+SN.}}
\end{figure*}

\begin{figure*}[!htp]
\includegraphics[width=0.45\textwidth]{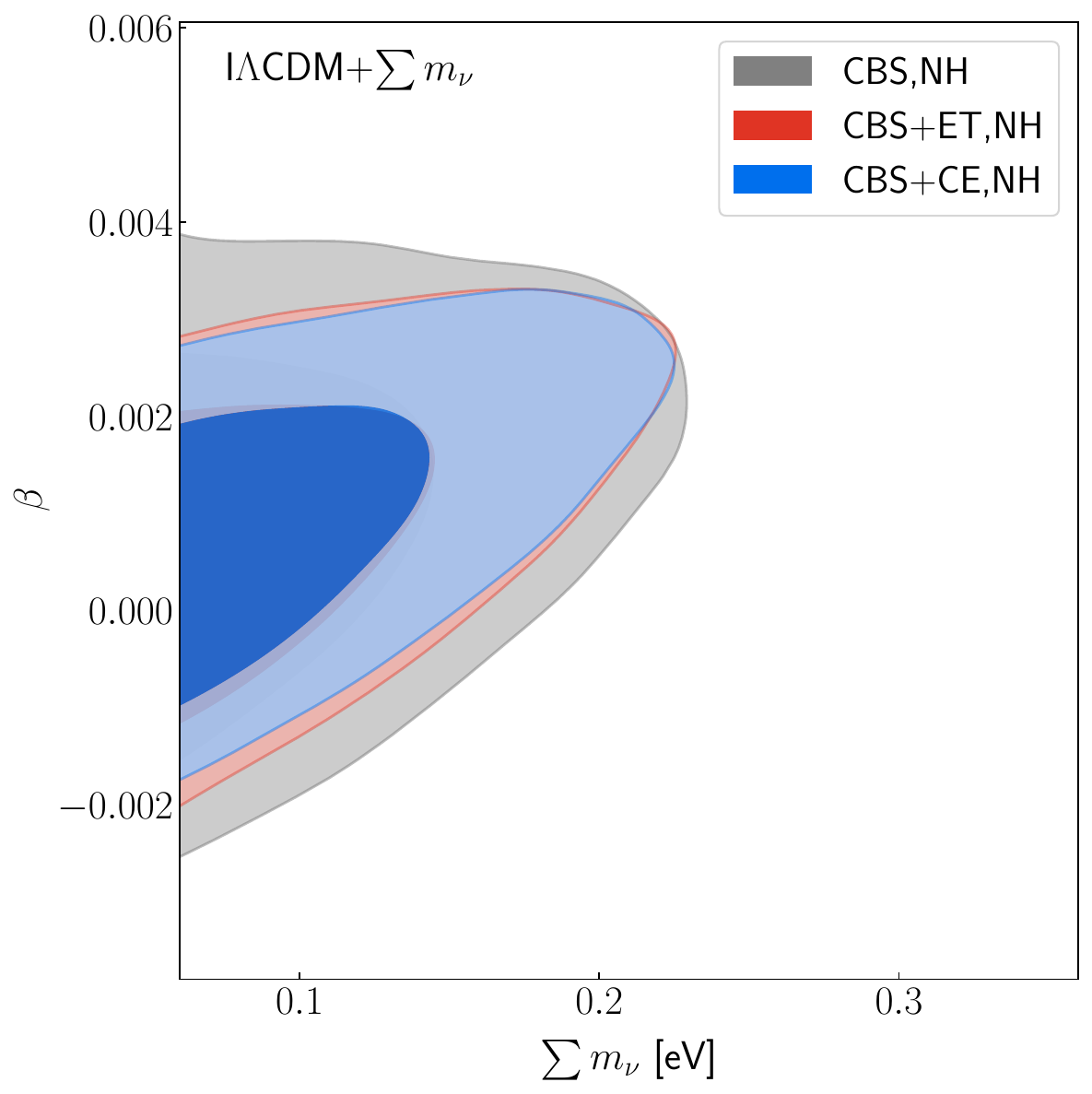}\ \hspace{1cm}
\includegraphics[width=0.45\textwidth]{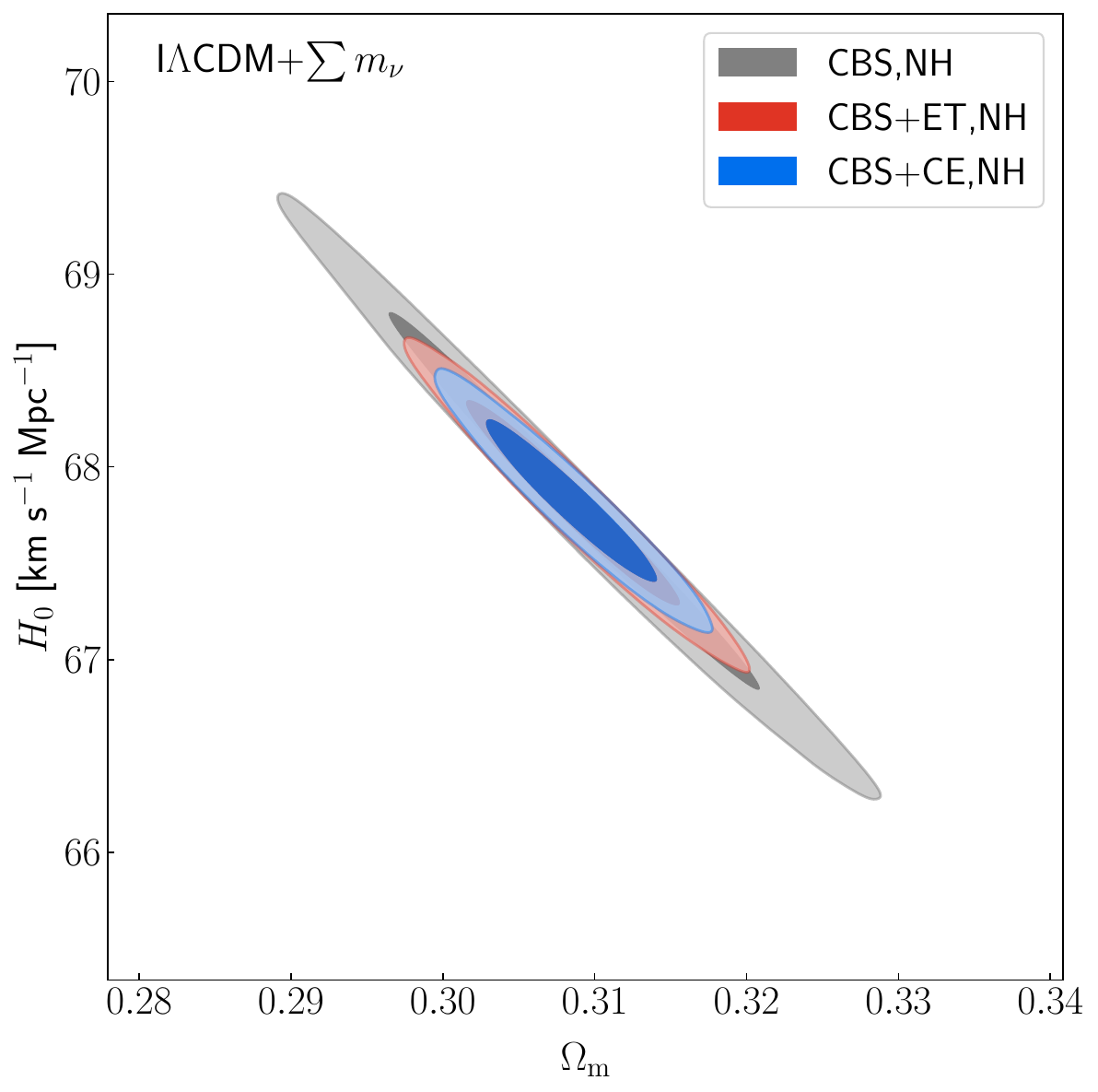}\ \hspace{1cm}
\centering \caption{\label{fig5} {Two-dimensional marginalized contours (68.3\% and 95.4\% confidence level) in the $\sum m_\nu$--$\beta$ and $\Omega_{\rm m}$--$H_0$ planes using the CBS, CBS+ET, and CBS+CE data. Here CBS stands for CMB+BAO+SN.}}
\end{figure*}

\begin{figure*}[!htp]
\includegraphics[width=0.45\textwidth]{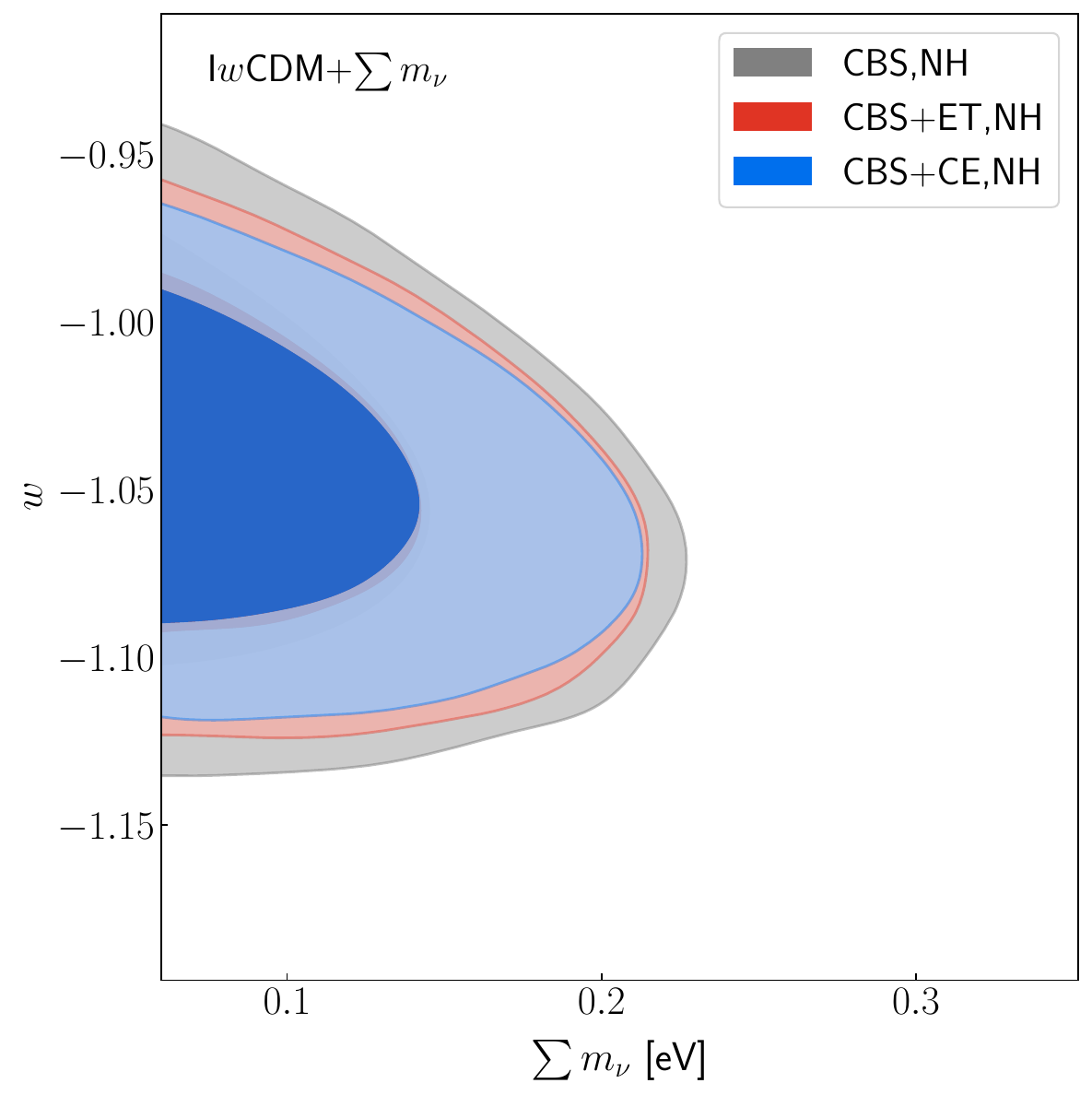}\ \hspace{1cm}
\includegraphics[width=0.45\textwidth]{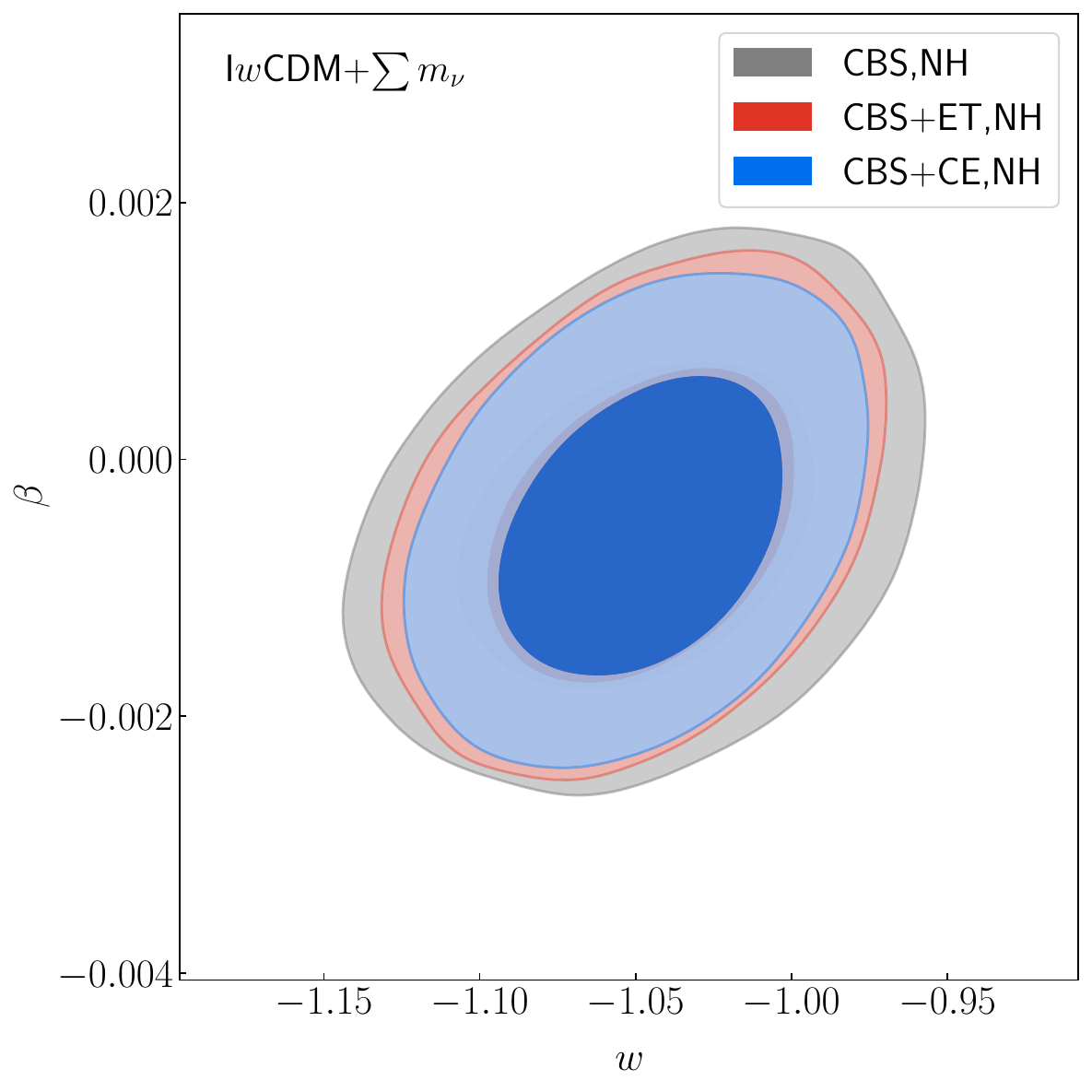}\ \hspace{1cm}
\centering \caption{\label{fig6} {Two-dimensional marginalized contours (68.3\% and 95.4\% confidence level) in the $\sum m_\nu$--$w$ and $w$--$\beta$ planes using the CBS, CBS+ET, and CBS+CE data. Here CBS stands for CMB+BAO+SN.}}
\end{figure*}

\begin{table*}[htbp]
	\caption{\label{tab1}{The absolute and relative errors of cosmological parameters in the $\Lambda$CDM+$\sum m_\nu$ model using the CBS, CBS+ET, and CBS+CE data. Note that $H_0$ is in units of km s$^{-1}$ Mpc$^{-1}$ and CBS stands for CMB+BAO+SN. Here, $2\sigma$ upper limits on $\sum m_\nu$ are given.}}
	\setlength\tabcolsep{0.5pt}
	\renewcommand{\arraystretch}{1.5}
	\centering
	{
		\begin{tabular}{cccccccccccccc}\\
			\hline
			\multicolumn{1}{c}{$\Lambda$CDM}&&\multicolumn{3}{c}{CMB+BAO+SN}&&\multicolumn{3}{c}{CMB+BAO+SN+ET}&&\multicolumn{3}{c}{CMB+BAO+SN+CE}\cr
			\cline{1-1}\cline{3-5}\cline{7-9}\cline{11-13}
			 Parameter&&NH&IH&DH&&NH&IH&DH&&NH&IH&DH\cr
			\hline
			$\sigma(\Omega_{\rm m})$&&$0.0062$&$0.0062$&$0.0064$&&$
			0.0044$&$0.0044$&$0.0043$&&$0.0036$&$0.0037$&$0.0036$\cr
			$\sigma(H_0)$&&$0.47$&$0.46$&$0.49$&&$0.32$&$0.32$&$
			0.32$&&$0.26$&$0.26$&$0.26$\cr
			
			$\ve(\Omega_{\rm m})$&&$1.98\%$&$1.97\%$&$2.07\%$
			&&$1.41\%$&$1.40\%$&$1.39\%$
			
			&&$1.15\%$&$1.18\%$&$1.16\%$\cr
			
			$\ve(H_0)$&&$0.70\%$&$0.68\%$&$0.72\%$
			
			&&$0.47\%$&$0.48\%$&$0.47\%$
			
			&&$0.39\%$&$0.39\%$&$0.38\%$\cr
			
			$\sum m_\nu$ [eV]&&$\ <0.156$&$\ <0.184$&$\ <0.121$&&$\ <0.146$&$\ <0.179$&$\ <0.106$&&$\ <0.144$&$\ <0.176$&$\ <0.104$\cr
			\hline
		\end{tabular}
	}
\end{table*}

\begin{table*}[htbp]
	\caption{\label{tab2} {The absolute and relative errors of cosmological parameters in the I$\Lambda$CDM+$\sum m_\nu$ model using the CBS, CBS+ET, and CBS+CE data. Note that $H_0$ is in units of km s$^{-1}$ Mpc$^{-1}$ and CBS stands for CMB+BAO+SN. Here, $2\sigma$ upper limits on $\sum m_\nu$ are given.}}
	\setlength\tabcolsep{0.5pt}
	\renewcommand{\arraystretch}{1.5}
	\centering
	{
		\begin{tabular}{cccccccccccccc}\\
			\hline
			\multicolumn{1}{c}{I$\Lambda$CDM}&&\multicolumn{3}{c}{CMB+BAO+SN}&&\multicolumn{3}{c}{CMB+BAO+SN+ET}&&\multicolumn{3}{c}{CMB+BAO+SN+CE}\cr
			\cline{1-1}\cline{3-5}\cline{7-9}\cline{11-13}
			Parameter&&NH&IH&DH&&NH&IH&DH&&NH&IH&DH\cr
			\hline
		    $\sigma(\Omega_{\rm m})$&&$0.0081$&$0.0082$&$0.0081$
			&&$0.0046$&$0.0045$&$0.0046$
			&&$0.0037$&$0.0037$&$0.0037$
			\cr
			$\sigma(H_0)$&&$0.65$&$0.65$&$0.65$
			&&$0.35$&$0.35$&$0.36$
			&&$0.27$&$0.27$&$0.27$
			\cr
			$\sigma(\beta)$&&$ 0.0013$&$0.0013$&$0.0013$
			&&$0.00104$&$0.00103$&$0.00105$
			&&$0.00096$&$ 0.00096$&$0.00101$
			\cr
			$\ve(\Omega_{\rm m})$
			&&$2.62\%$&$2.65\%$&$2.65\%$
			&&$1.49\%$&$1.46\%$&$1.50\%$
			&&$1.20\%$&$1.20\%$&$1.20\%$
			\cr
			$\ve(H_0)$
			&&$0.96\%$&$0.96\%$&$0.96\%$
			&&$0.52\%$&$0.52\%$&$0.53\%$
			&&$0.40\%$&$0.40\%$&$0.40\%$
			\cr
			$\sum m_\nu$ [eV]
			&&$< 0.191$&$\ < 0.224$&$\ < 0.148$
			&&$\ < 0.188$&$\ < 0.220$&$\ < 0.147$
			&&$\ < 0.187$&$\ < 0.220$&$\ < 0.142$\cr
			\hline
		\end{tabular}
	}
\end{table*}

\begin{table*}[htbp]
	\caption{\label{tab3}{The absolute and relative errors of cosmological parameters in the I$w$CDM+$\sum m_\nu$ model using the CBS, CBS+ET, and CBS+CE data. Note that $H_0$ is in units of km s$^{-1}$ Mpc$^{-1}$ and CBS stands for CMB+BAO+SN. Here, $2\sigma$ upper limits on $\sum m_\nu$ are given.}}
	\setlength\tabcolsep{0.5pt}
	\renewcommand{\arraystretch}{1.5}
	\centering
	{
		\begin{tabular}{cccccccccccccc}\\
			\hline
			\multicolumn{1}{c}{I$w$CDM}&&\multicolumn{3}{c}{CMB+BAO+SN}&&\multicolumn{3}{c}{CMB+BAO+SN+ET}&&\multicolumn{3}{c}{CMB+BAO+SN+CE}\cr
			\cline{1-1}\cline{3-5}\cline{7-9}\cline{11-13}
			Parameter&&NH&IH&DH&&NH&IH&DH&&NH&IH&DH\cr
			\hline
			$\sigma(\Omega_{\rm m})$&&$0.0079$&$0.0078$&$0.0079$&&$
			0.0048$&$0.0048$&$0.0048$
			&&$0.0039$&$0.0039$&$0.0039$\cr
			
			$\sigma(H_0)$&&$0.82$&$0.81$&$0.82$
			&&$0.54$&$0.54$&$0.54$
			&&$0.46$&$0.46$&$0.45$\cr
			
			$\sigma(w)$&&$0.037$&$0.037$&$0.038$
			&&$0.033$&$0.033$&$0.033$
			&&$0.030$&$0.030$&$0.030$\cr
			
			$\sigma(\beta)$&&$0.00085$&$0.00087$&$0.00088$
			&&$0.00083$&$0.00080$&$0.00082$
			&&$0.00079$&$0.00079$&$0.00079$\cr
			
			$\ve(\Omega_{\rm m})$
			&&$2.56\%$&$2.52\%$&$2.57\%$
			&&$1.56\%$&$1.55\%$&$1.55\%$
			&&$1.27\%$&$1.26\%$&$1.27\%$\cr
			
			$\ve(H_0)$
			&&$1.20\%$&$1.19\%$&$1.20\%$
			&&$0.79\%$&$0.79\%$&$0.79\%$
		    &&$0.67\%$&$0.67\%$&$0.66\%$\cr
			
			$\ve(w)$
			&&$3.52\%$&$3.50\%$&$3.56\%$
			&&$3.15\%$&$3.13\%$&$3.13\%$
			&&$2.90\%$&$2.84\%$&$2.89\%$\cr

			$\sum m_\nu$ [eV]&&$< 0.190$&$\ < 0.224$&$\ < 0.149$
			&&$\ < 0.182$&$\ < 0.212$&$\ < 0.146$
			&&$\ < 0.180$&$\ < 0.210$&$
			 <0.136$\cr
			\hline
		\end{tabular}
	}
\end{table*}

\section{Results and discussion}\label{sec3}
In this section, we report the constraint results of cosmological parameters in the $\Lambda$CDM+$\sum m_{\nu}$, I$\Lambda$CDM+$\sum m_{\nu}$, and I$w$CDM+$\sum m_{\nu}$ models. In these models, the three mass hierarchy cases of neutrinos, i.e., the NH, IH, and DH cases, have been considered. The constraint results of the NH case are shown as a representative in Figs.~\ref{fig4}--\ref{fig6} and the constraint results are summarized in Tables \ref{tab1}--\ref{tab3}. Note that for the constraints on the total neutrino mass, the $2\sigma$ upper limits are given. Note also that using the squared mass differences derived from the neutrino oscillation experiments, one can obtain the lower limits for the total neutrino mass, i.e., 0.05 eV for NH and 0.1 eV for IH; in the case of DH, the smallest value of the total neutrino mass is zero.
For a parameter $\xi$, we use $\sigma(\xi)$ and $\varepsilon(\xi)$ to represent its absolute and relative errors, respectively, with $\varepsilon(\xi)$ defined as $\varepsilon(\xi)=\sigma(\xi)/\xi$.

We first take a look at the results in the $\Lambda$CDM+$\sum m_{\nu}$ model. In Fig.~\ref{fig4}, we show the constraints on the $\Lambda$CDM+$\sum m_{\nu}$ model in the $\sum m_{\nu}$--$H_0$ and $\Omega_{\rm m}$--$H_0$ planes from the CBS, CBS+ET, and CBS+CE data. We find that the addition of the GW data to the CBS data could lead to the reduction of the upper limits of $\sum m_{\nu}$ to some extent. The CBS+CE data give slightly smaller upper limits on $\sum m_{\nu}$ than those from the CBS+ET data. Concretely, when adding the ET data to the CBS data, the upper limits on $\sum m_{\nu}$ could be reduced by 2.7\%--12.4\% in the three hierarchy cases. While for CE, the upper limits on $\sum m_{\nu}$ could be reduced by 4.3\%--14.0\% in the three hierarchy cases. Here the results of ET are consistent with the previous results in Ref.~\cite{Wang:2018lun}.

Although using the GW data could only slightly improve the limits on the neutrino mass, they can significantly help improve the constraints on other cosmological parameters.
We find that the constraints on $\Omega_{\rm m}$ and $H_0$ could be improved by 29.0\%--32.8\% and 30.4\%--34.7\%, respectively, when adding the ET data to the CBS data, and by 40.3\%--43.8\% and 43.5\%--46.9\%, respectively, for the case of CE.


In Fig.~\ref{fig5}, we show the constraints on the I$\Lambda$CDM+$\sum m_{\nu}$ model in the $\sum m_{\nu}$--$\beta$ and $\Omega_{\rm m}$--$H_0$ planes from the CBS, CBS+ET, and CBS+CE data. We can clearly see that, when considering the interaction between vacuum energy and dark matter, the improvement of the limits on $\sum m_\nu$ by adding GW data is rather not evident. In the case of ET, the improvement of the limit on $\sum m_\nu$ is only 0.7\%--1.8\%, and in the case of CE, the improvement is 1.8\%--4.1\%. Therefore, we find that compared with the standard $\Lambda$CDM model, in its interaction version, the I$\Lambda$CDM model, the improvement of the limits on $\sum m_\nu$ by GW data from ET and CE becomes weaker. {This is because the I$\Lambda$CDM model considers an extra cosmological parameter $\beta$ compared with the $\Lambda$CDM model, which will degenerate with other cosmological parameters when the CBS data are used to constrain the I$\Lambda$CDM model. Hence, compared with the $\Lambda$CDM model, the addition of the GW data to the CBS data for its interaction version leads to weaker improvement.}

We also find that the constraints on the coupling parameter $\beta$ can be improved by using the GW data to a certain extent. In the I$\Lambda$CDM+$\sum m_{\nu}$ model, the constraints on $\beta$ are improved by 19.2\%--20.8\% and 22.3\%--26.2\%, respectively, when the GW data of ET and CE are added on the basis of the CBS case.


In Fig.~\ref{fig6}, we show the constraints on the I$w$CDM+$\sum m_{\nu}$ model in the $\sum m_{\nu}$--$w$ and $w$--$\beta$ planes from the CBS, CBS+ET, and CBS+CE data. We find that in this case the improvement of the limits on the neutrino mass is better than in the previous case. {For ET, the improvement of the limit on $\sum m_\nu$ is 2.0\%--5.4\%, and for CE, the improvement is 5.3\%--8.7\%.}

We find that in this case the constraints on the coupling parameter $\beta$ and the EoS parameter of dark energy $w$ can both be significantly improved by considering the addition of GW data. {The constraints on $\beta$ and $w$ are improved by 2.4\%--8.0\% and 10.8\%--13.2\%, respectively, when considering the ET data, and by 7.1\%--10.2\% and 18.9\%--21.1\%, respectively, when considering the CE data.}

In this work, we discuss the cosmological constraints on the IDE models in the cases of considering the GW standard siren observations from 3G ground-based GW detectors ET and CE. The results show that the limits on the neutrino mass can only be slightly improved with the help of the GW data, on the basis of the CBS constraint. Since the GW data can precisely constrain the Hubble constant $H_0$, the addition of them in the cosmological fit can help break the cosmological parameter degeneracies formed by other cosmological observations. Therefore, the consideration of GW standard siren data can help significantly improve the constraints on the most cosmological parameters. However, the effect of massive neutrinos in the late universe and on the large scales cannot be distinctively distinguished from that of the cold dark matter, leading to the improvement of the limits on the neutrino mass by considering GW data is not obvious. Anyway, even though the impact on constraining the neutrino mass is not apparent, the GW standard sirens are rather useful in helping improve the constraints on the most cosmological parameters including the EoS of dark energy and the coupling between dark energy and dark matter.

\section{Conclusion}\label{sec4}

In the era of 3G ground-based GW detectors, a lot of GW standard siren data with the know redshifts could be obtained by the multi-messenger observation for BNS merger events. Obviously, these standard sirens would exert great impacts on the cosmological parameter estimation. Since the GW standard sirens can tightly constrain the Hubble constant, the consideration of them in a joint cosmological fit can lead to the cosmological parameter degeneracies formed by other cosmological observations being well broken. The GW standard sirens can thus be used to help significantly improve the constraints on cosmological parameters in the future.

It is of great interest to investigate whether the limits on the total neutrino mass can also be effectively improved by considering the GW standard siren data. In particular, the cosmological constraints on the neutrino mass are strongly model-dependent, and so the cases in different cosmological scenarios are needed to be detailedly discussed. In this work, we discuss the issue of weighing neutrinos in the IDE models by using the GW standard siren observations by ET and CE.

We consider the simplest IDE models, namely the I$\Lambda$CDM and I$w$CDM models with $Q=\beta H \rho_{\mathrm{c}}$. We simulate the GW standard siren data of the BNS mergers observed by ET and CE (in a way of multi-messenger detection). We investigate whether the GW standard sirens observed by ET and CE could help improve the constraint on the neutrino mass in the IDE models.

It is found that the GW standard siren observations from ET and CE can only slightly improve the constraint on the neutrino mass in the IDE models, compared to the current limit given by CMB+BAO+SN.
This is mainly because the effect of massive neutrinos in the late universe and on the large scales cannot be distinctively distinguished from that of the CDM, leading to the improvement of the limits on the neutrino mass by considering GW data is not obvious.
Although the limit on neutrino mass can only be slightly updated by considering the GW standard sirens, they are fairly useful in helping improve the constraints on the most cosmological parameters including the EoS of dark energy and the coupling between dark energy and dark matter.

\begin{acknowledgments}
This work was supported by the National Natural Science Foundation of China (Grants Nos. 11975072, 11835009, 11875102, and 11690021), the Liaoning Revitalization Talents Program (Grant No. XLYC1905011), the Fundamental Research Funds for the Central Universities (Grant No. N2005030), the National 111 Project of China (Grant No. B16009), and the Science Research Grants from the China Manned Space Project (Grant No. CMS-CSST-2021-B01).

\end{acknowledgments}

\bibliography{idegw}

\end{document}